\documentclass[conference]{IEEEtran}
\IEEEoverridecommandlockouts
\usepackage{algorithm,algpseudocode,amsmath,amssymb,amsfonts,array,ifthen,mathrsfs,physics,tabularx,tikz,circuitikz,tikz-dsp,pgf,siunitx}
\usepackage[nolist,nohyperlinks]{acronym}
\usetikzlibrary{shapes.misc,shapes.geometric,positioning,fit,backgrounds,patterns,shapes.arrows,fadings}
\usepackage[utf8]{inputenc}
\DeclareUnicodeCharacter{2212}{−}
\newlength\figH
\newlength\figW
\newlength\figHhalf
\newlength\figWhalf
\DeclareMathOperator*{\argmin}{\arg\!\min}

\everymath=\expandafter{\the\everymath\displaystyle}

\newcommand{\E}[1]{\mathop{\mathbb{E}} \left[ #1 \right]}
\newcommand{\EsG}{\hat{\mathbf{G}}}
\newcommand{\EsGLS}{\EsG_\mathrm{LS}}

\newcommand{\Esp}{\hat{\mathbf{p}}}
\newcommand{\EG}{\mathbf{E}_{\EsG}}
\newcommand{\EGLS}{\mathbf{E}_{\EsGLS}}
\newcommand{\EX}{\mathbf{E}_{\mathbf{X}}}
\newcommand{\EZ}{\mathbf{E}_{\mathbf{Z}}}

\newcommand{\GG}{\mathbf{G}}

\newcommand{\HH}{\mathsf{H}}
\newcommand{\II}{\mathbf{I}}

\newcommand{\jj}{\mathrm{j}}

\newcommand{\lsb}{\Delta_{b}}
\newcommand{\lsbrx}{\Delta_{b_\mathrm{Rx}}}
\newcommand{\lsbtx}{\Delta_{b_\mathrm{Tx}}}

\newcommand{\NN}{\mathbf{N}}

\renewcommand{\Tr}[1]{\mathop{\mathrm{Tr}} \! \left\lbrace #1 \right\rbrace}
\newcommand{\Vmax}{V_\mathrm{Max}}

\newcommand{\WW}{\mathbf{W}}
\newcommand{\XX}{\mathbf{X}}
\newcommand{\YY}{\mathbf{Y}}
\newcommand{\ZZ}{\mathbf{Z}}

\begin{document}

\bstctlcite{IEEEexample:BSTcontrol}

\begin{acronym}
	\acro{6G}[6G]{sixth generation of mobile communications standard}
	\acro{ADC}[ADC]{analog-to-digital converter}
    \acro{DAC}[DAC]{digital-to-analog converter}
    \acro{HER}[HermesPy]{heterogeneous radio mobile simulator Python}
    \acro{IQ}[IQ]{in-phase / quadrature}
    \acro{LSB}[LSB]{least significant bit}
    \acro{MIMO}[MIMO]{multiple-input multiple-output}
    \acro{RF}[RF]{radio-frequency}
    \acro{SDR}[SDR]{software defined radio}
	\acro{ULA}[ULA]{uniform linear array}
    \acro{USRP}[USRP]{universal software radio peripheral}
\end{acronym}

\author{%
\IEEEauthorblockN{
        Jan Adler\IEEEauthorrefmark{1},
        Florian Gast\IEEEauthorrefmark{1},
        Gerhard P. Fettweis\IEEEauthorrefmark{1}\IEEEauthorrefmark{2},
        and Rafael F. Schaefer\IEEEauthorrefmark{1}\IEEEauthorrefmark{2}\\[1ex]
    }
    \IEEEauthorblockA{
        \IEEEauthorrefmark{1}Barkhausen Institut, Dresden, Germany\\
   }
    \IEEEauthorblockA{
        \IEEEauthorrefmark{2}Technische Universität Dresden, Dresden, Germany\\[0.5ex]
   }
    \IEEEauthorblockA{
        \{jan.adler, florian.gast, gerhard.fettweis, rafael.schaefer\}@barkhauseninstitut.org}
\thanks{%
This work was financed with tax revenue on the basis of the budget approved by the Saxon state parliament, the European Union’s Horizon Europe SNS project INSTINCT (grant 101139161), the German Research Foundation's Cluster of Excellence CeTI (project 390696704) and the German Federal Ministry of Research, Technology and Space through the transfer hub 6G-life (grant 16KIS2413K).
}%
}

\title{Quantization Limitations of Leakage Suppression in Self-Calibrating Monostatic Integrated Sensing and Communication MIMO Systems}

\maketitle

\begin{abstract}
Power leaking directly from transmitting into receiving radio-frequency chains is a key challenge in the realization of monostatic sensing applications with multi-antenna communication front-ends, to which a promising solution is digitally precoding transmitted signals for improved leakage suppression.
While digital transmit precodings perform well in theory, real-world deployments typically exhibit severely degraded leakage suppression.
This work investigates quantization noise as a primary factor limiting the performance of such precoding schemes.
A closed-form solution predicting the impact of quantization noise on the performance of arbitrary digital joint leakage estimation and leakage suppression precodings is derived, numerically analyzed, and validated in a hardware testbed.
\end{abstract}

\section{Introduction} \label{sec:intro}
The integration of sensing functionalities into existing communication devices and infrastructure is one of the novel features to be adopted with the upcoming \ac{6G} \cite{wymeersch.2014,liu.2022}.
Sensing applications with future \ac{6G} and existing communication \ac{RF} \ac{MIMO} front-ends can generally be realized by implementing a multistatic or monostatic radar architecture \cite{liu.2022}.
While multistatic sensing estimates environmental parameters from the waveform propagation between two spatially separated devices, monostatic sensing estimates environmental parameters from the waveform backscattered to receiving antennas closely co-located to transmitting antennas within a single device.
Both architectures offer unique advantages and drawbacks:
Multistatic sensing, on the one hand, is limited to reference and synchronization sections of the deployed waveform and suffers from carrier frequency offsets and timing synchronization mismatches, resulting in position and velocity ambiguities of the estimated sensing parameters.
Monostatic sensing, on the other hand, like full-duplex communication, suffers from the transmitted power directly leaking from the transmit \ac{RF} chains and antennas into the receiving \ac{RF} chains and antennas, driving the receiving amplifiers into saturation and clipping the receiving \acp{ADC} \cite{Sabharwal.2014,hong.2014,zhang.2022}.
It should be noted that, while full-duplex communication aims to suppress the full self-interference, including direct leakage as well as all backscattered power from the environment, in order to minimize interference with incoming communication waveforms, basic sensing applications often require only a suppression of the direct transmit-receive leakage, since the environmental backscattering's parameters are exactly the parameters of interest to be estimated.
Leakage may be addressed by designing specialized antenna arrays with minimized coupling between the transmitting and receiving antenna elements \cite{yalcinkaya.2023} and by optimizing \ac{RF} front-ends \cite{bozorgi.2021}.
Attempts have been made to cancel out leakage in the \ac{RF} domain by either a physical loopback between individual transmit and receive \ac{RF} chains \cite{riihonen.2012} or by deploying a dedicated \ac{DAC} \cite{zhang.2022}.
These techniques, however, naturally scale poorly with an increasing number of transmitting and receiving antennas.
Instead, a promising approach within monostatic \ac{MIMO} arrays is optimizing the transmit beamformer for improved leakage suppression \cite{everett.2016,adler.2025a,hernangomez.2026}.
Digital transmit-beamforming leakage suppression performs nearly perfectly in simulations with overly idealistic hardware assumptions, suppressing the remaining leakage power well below the noise floor.
In real-world demonstrations, however, the expected leakage reduction is observable, but limited \cite{everett.2016,adler.2025a}.
The primary hypothesis of this work is the proposition that quantization noise is a major, if not the primary, contributor to limiting the performance of leakage-minimizing digital transmit precoding schemes.
The presented considerations are structured as follows:
Section \ref{sec:sys} introduces a simplified linear system model of a quantized monostatic integrated sensing and communication \ac{MIMO} architecture, based on which Section \ref{sec:quantization} derives a closed-form expression for the expected leakage suppression performance.
Sections \ref{sec:simulation} and \ref{sec:measurements} validate the derived performance expectations by Monte Carlo simulations and a measurement campaign in a \ac{SDR} \ac{MIMO} testbed, respectively.
The argument is concluded with Section \ref{sec:conclusion}.
Python implementations to reproduce the presented results may be accessed through a public GitHub repository~\cite{adler.2026.quantization_repository}.

\section{System Model} \label{sec:sys}

The quantization-effects of a uniform mid-tread \ac{ADC} or uniform mid-tread \ac{DAC} with $b$ bits on the amplitude of an incoming signal $x^{(k)} \in \mathbb{R}$ at discrete time instance $k$ can be modeled as the operation
\begin{equation}
	U_b(x^{(k)}) = \lsb \left\lceil \frac{x^{(k)}}{\lsb} - \frac{1}{2} \right\rceil \in \mathbb{U}_b = \lbrace u_{b}^{(1)}\dotsc u_{b}^{(2^{b})}  \rbrace \; \text {,}
\end{equation}
rounding from continuous-value analog to discrete-value digital domain and vice-versa to a set of $2^b$ discrete signal levels $\mathbb{U}_b$.
For improved readability, the time-discrete notation over index $k$ will be maintained for signals in the time-continuous domain between \acp{DAC} and \acp{ADC}.
A maximum quantization range of $\Vmax > 0$ translates to a quantization step size, represented by the \ac{LSB}, of
\begin{equation}
	\lsb = \frac{2\Vmax}{2^b-1}
\end{equation}
and subsequently to the discrete quantization levels
\begin{equation}
    u_{b}^{(q)} = -\Vmax + \big(q - \frac{1}{2}\big) \lsb \text{ .}
\end{equation}
Assuming the quantizer's input remains within non-clipping input range, i.e., $|x| \leq \Vmax$, its introduced error $\epsilon_b$ can be modeled as a uniformly distributed noise, i.e., the ADC output can be described as
\begin{equation}
	U_b(x^{(k)}) = x^{(k)} + \epsilon_b^{(k)} \quad \text{with} \quad \epsilon_b^{(k)} \sim \mathcal{U}\Big(-\frac{\lsb}{2}, \frac{\lsb}{2}\Big)
\end{equation}
added to the value-continuous input $x$, cf. \cite{katzenelson.1962}.
This model can be extended to complex numbers $x^{(k)} = x^{(k)}_\mathrm{Re} + \jj x^{(k)}_\mathrm{Im} \in \mathbb{C}$ by treating real part $x^{(k)}_\mathrm{Re}$ and imaginary part $x^{(k)}_\mathrm{Im}$ independently and denoting the imaginary unit by $\jj = \sqrt{-1}$, so that 
\begin{align}
    \begin{split}
    \widetilde{U}_b(x^{(k)}) &= U_b(x^{(k)}_\mathrm{Re}) + \jj U_b(x^{(k)}_\mathrm{Im}) \in \widetilde{\mathbb{U}}_b \\
    &= x^{(k)} + \tilde{\epsilon}^{(k)}_b \; \text{with } \; \tilde{\epsilon}^{(k)}_b \sim \mathcal{CU}\Big(-\frac{\lsb}{2}, \frac{\lsb}{2}\Big) \; \text{,}
    \end{split}
\end{align}
introducing complex noise independently and uniformly distributed in the real and imaginary parts, and mapping to the discrete complex set
\begin{equation}
    \widetilde{\mathbb{U}}_b = \lbrace a + \jj b : a, b \in \mathbb{U}_b \rbrace \; \text{.}
\end{equation}
By adopting the notation
\begin{equation}
   m^{(a,k)} = m_\mathrm{Re}^{(a,k)} + \jj m_\mathrm{Im}^{(a,k)} \in \mathbb{C} \quad \text{in} \quad \mathbf{M} \in \mathbb{C}^{A \times K}
\end{equation}
as the element in the $a\mathrm{-th}$ row and $k\mathrm{-th}$ column of an arbitrary complex-valued matrix $\mathbf{M}$ of height $A$ and width $K$, the quantization operation may furthermore be extended to matrices as
\begin{equation}
    \widetilde{U}_b(\mathbf{M}) = \widetilde{\mathbf{M}} = \mathbf{M} + \mathbf{E}_b \quad \text{with} \quad \tilde{m}^{(a,k)} = \widetilde{U}_b(m^{(a,k)})
\end{equation}
introducing a uniformly distributed error matrix $\mathbf{E}_b$ with mean and variance
\begin{equation}
    \E{\mathbf{E}_b} = 0, \quad \E{\mathbf{E}_b^\HH \mathbf{E}_b} = A\frac{\lsb^2}{6}\II_K \; \text{.}
\end{equation}%
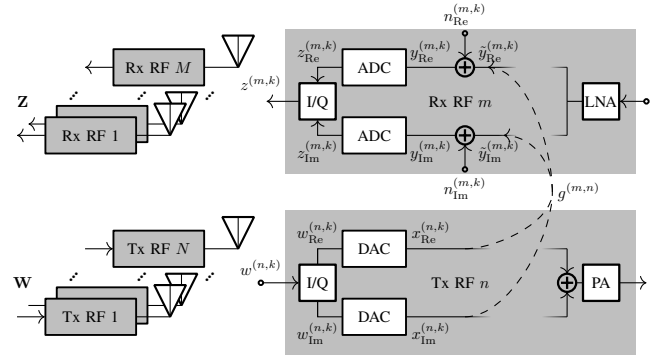
\begin{figure}
	\centering
	\begin{tikzpicture}[scale=0.6, every node/.style={scale=0.6}]

    \coordinate (anchor_compressed_tx) at (0.0,0.0) {};
    \coordinate (anchor_compressed_rx) at (0.0,4.0) {};
    \coordinate (anchor_tx) at (6.5,0.0) {};
    \coordinate (anchor_rx) at (6.5,4.0) {};

    \tikzstyle{dots} = [rotate=45.5, behind path]

    \foreach \n in {6, 1, 0} {
        \pgfmathsetmacro{\rflabelindex}{ifthenelse(\n==6,"N","$\the\numexpr\n+1\relax$")}

        \node (tx_comp_in_\n) at ($(anchor_compressed_tx)+(0.25*\n,0.25*\n)$) [] {};
        \node (tx_comp_rf_in_\n) at ($(tx_comp_in_\n)+(1.75,0.0)$) [dspfilter,fill=lightgray,minimum width=2cm] {Tx RF $\rflabelindex$};
        \node (tx_comp_ant_\n) at ($(tx_comp_rf_in_\n)+(1.75,0.0)$) [bareantenna, fill=white] {};

        \draw[->] (tx_comp_in_\n) -- (tx_comp_rf_in_\n) {};
        \draw (tx_comp_rf_in_\n) -- (tx_comp_ant_\n) {};
    }

    \node at ($(tx_comp_in_0)!0.5!(tx_comp_in_6)-(0.5,0.0)$) {$\mathbf{W}$};

    \foreach \n in {6, 1, 0} {
        \pgfmathsetmacro{\rflabelindex}{ifthenelse(\n==6,"M","$\the\numexpr\n+1\relax$")}

        \node (rx_comp_in_\n) at ($(anchor_compressed_rx)+(0.25*\n,0.25*\n)$) [] {};
        \node (rx_comp_rf_in_\n) at ($(rx_comp_in_\n)+(1.75,0.0)$) [dspfilter,fill=lightgray,minimum width=2cm] {Rx RF $\rflabelindex$};
        \node (rx_comp_ant_\n) at ($(rx_comp_rf_in_\n)+(1.75,0.0)$) [bareantenna, fill=white] {};

        \draw[->] (rx_comp_rf_in_\n) -- (rx_comp_in_\n) {};
        \draw (rx_comp_ant_\n) -- (rx_comp_rf_in_\n) {};
    }

    \node at ($(rx_comp_in_0)!0.5!(rx_comp_in_6)-(0.5,0.0)$) {$\ZZ$};
    
    \foreach \txrx in {tx, rx} {
        \node [dots] at ($(\txrx_comp_in_1)!0.5!(\txrx_comp_in_6)+(0.5,0.0)$) {\textbf{...}};
        \node [dots] at ($(\txrx_comp_rf_in_1)!0.5!(\txrx_comp_rf_in_6)+(0.25,0.0)$) {\textbf{...}};
        \node [dots] at ($(\txrx_comp_ant_1)!0.5!(\txrx_comp_ant_6)$) {\textbf{...}};
    }

    \node [label=above:{$w^{(n,k)}$}]  (tx_digital) at ($(anchor_tx)+(-1, 0.75)$) [dspnodeopen] {};

    \node (splitter_tx) at ($(tx_digital)+(1.25,0.0)$) [dspsquare, fill=white] {I/Q};
    \coordinate [label=above:{$w_\mathrm{Re}^{(n,k)}$}] (splitter_tx_edge_i) at ($(splitter_tx)+(0.0,0.75)$) {};
    \coordinate [label=below:{$w_\mathrm{Im}^{(n,k)}$}] (splitter_tx_edge_q) at ($(splitter_tx)-(0.0,0.75)$) {};

    \node (dac_tx_i) at ($(anchor_tx)+(1.5,1.5)$) [dspfilter, fill=white] {DAC};
    \node (dac_tx_q) at ($(anchor_tx)+(1.5,0.0)$) [dspfilter, fill=white] {DAC};
    \coordinate [label=above:{$x_\mathrm{Re}^{(n,k)}$}] (xxx) at ($(anchor_tx)+(2.75,1.5)$) {};
    \coordinate [label=below:{$x_\mathrm{Im}^{(n,k)}$}] (xxx) at ($(anchor_tx)+(2.75,0.0)$) {};

    \coordinate (tx_anchor_g_i) at ($(dac_tx_i)+(2.0,0.0)$) {};
    \coordinate (tx_anchor_g_q) at ($(dac_tx_q)+(2.0,0.0)$) {};



    \coordinate (adder_tx_edge_i) at ($(tx_anchor_g_i)+(2.25,0.0)$) {};
    \coordinate (adder_tx_edge_q) at ($(tx_anchor_g_q)+(2.25,0.0)$) {};
    \node (adder_tx) at ($(adder_tx_edge_i)!0.5!(adder_tx_edge_q)$) [dspadder, fill=white] {};

    \node (amp_tx) at ($(adder_tx)+(0.75,0.0)$) [dspsquare, fill=white] {PA};

    \coordinate (ant_tx_port) at ($(amp_tx)+(1.0,0.0)$) {};

    \draw[->] (tx_digital) -- (splitter_tx) {};
    \foreach \p in {i, q} {
        \draw (splitter_tx) -- (splitter_tx_edge_\p) {};
        \draw (splitter_tx) -- (splitter_tx_edge_\p) -- (dac_tx_\p) -- (tx_anchor_g_\p) {};
        \draw[-,path fading=west] (adder_tx_edge_\p) -- ($(tx_anchor_g_\p)!0.75!(adder_tx_edge_\p)$) {};
        \draw[-,path fading=east] (tx_anchor_g_\p) -- ($(tx_anchor_g_\p)!0.25!(adder_tx_edge_\p)$) {};
        \draw[->] (adder_tx_edge_\p) -- (adder_tx) {};
    }
    \draw[-] (adder_tx) -- (amp_tx) {};
    \draw[->] (amp_tx) -- (ant_tx_port) {};

    \node [label=above:{$z^{(m,k)}$}] (rx_digital) at ($(anchor_rx)+(-1,0.75)$) [] {};

    \node (combiner_rx) at ($(rx_digital)+(1.25,0.0)$) [dspsquare, fill=white] {I/Q};
    \coordinate [label=above:{$z_\mathrm{Re}^{(m,k)}$}] (combiner_rx_edge_i) at ($(combiner_rx)+(0.0,0.75)$) {};
    \coordinate [label=below:{$z_\mathrm{Im}^{(m,k)}$}] (combiner_rx_edge_q) at ($(combiner_rx)-(0.0,0.75)$) {};
    
    \node (adc_tx_i) at ($(anchor_rx)+(1.5,1.5)$) [dspfilter, fill=white] {ADC};
    \node (adc_tx_q) at ($(anchor_rx)+(1.5,0.0)$) [dspfilter, fill=white] {ADC};
    \coordinate [label=above:{$y_\mathrm{Re}^{(m,k)}$}] (y_i) at ($(anchor_rx)+(2.75,1.5)$) {};
    \coordinate [label=below:{$y_\mathrm{Im}^{(m,k)}$}] (y_q) at ($(anchor_rx)+(2.75,0.0)$) {};

    \node [label=above:{$n_{\mathrm{Re}}^{(m,k)}$}] (noise_source_rx_i) at ($(anchor_rx)+(3.5,2.25)$) [dspnodeopen, behind path, fill=white] {};
    \node [label=below:{$n_{\mathrm{Im}}^{(m,k)}$}] (noise_source_rx_q) at ($(anchor_rx)+(3.5,-0.75)$) [dspnodeopen, behind path, fill=white] {};
    \node (noise_sum_rx_i) at ($(anchor_rx)+(3.5,1.5)$) [dspadder,fill=white] {};
    \node (noise_sum_rx_q) at ($(anchor_rx)+(3.5,0.0)$) [dspadder,fill=white] {};

    \coordinate [label=above:{$\tilde{y}_\mathrm{Re}^{(m,k)}$}] (ytil_i) at ($(anchor_rx)+(4.25,1.5)$) {};
    \coordinate [label=below:{$\tilde{y}_\mathrm{Im}^{(m,k)}$}] (ytil_q) at ($(anchor_rx)+(4.25,0.0)$) {};

    \coordinate (rx_anchor_g_i) at ($(noise_sum_rx_i)+(0.5,0.0)$) {};
    \coordinate (rx_anchor_g_q) at ($(noise_sum_rx_q)+(0.85,0.0)$) {};

    \node (amp_rx) at ($(anchor_rx)+(6.5,0.75)$) [dspsquare,fill=white] {LNA};

    \coordinate (splitter_rx) at ($(amp_rx)-(0.75,0.0)$) {};
    \coordinate (splitter_rx_edge_i) at ($(splitter_rx)+(0.0,0.75)$) {};
    \coordinate (splitter_rx_edge_q) at ($(splitter_rx)+(0.0,-0.75)$) {};

    \node (ant_rx_port) at ($(amp_rx)+(1.0,0.0)$) [dspnodeopen,fill=white] {};

    \draw[->] (ant_rx_port) -- (amp_rx) {};
    \draw (amp_rx) -- (splitter_rx) {};
    \foreach \p in {i,q} {
        \draw[-] (splitter_rx) -- (splitter_rx_edge_\p) {};
        \draw[-,path fading=west] (splitter_rx_edge_\p) -- ($(ytil_\p)!0.75!(splitter_rx_edge_\p)$) {};
        \draw[-,path fading=east] (ytil_\p) -- ($(ytil_\p)!0.25!(splitter_rx_edge_\p)$) {};
        \draw[->] (ytil_\p) -- (noise_sum_rx_\p) -- (y_\p) -- (adc_tx_\p) -- (combiner_rx_edge_\p) -- (combiner_rx) {};
        \draw[->] (noise_source_rx_\p) -- (noise_sum_rx_\p) {};
    }
    \draw[->] (combiner_rx) -- (rx_digital) {};

    \begin{pgfonlayer}{background}
        \node[fit={($(tx_digital)-(0.0,1.7)$) ($(ant_tx_port)+(0.3,1.7)$)},fill=lightgray,behind path,scale=1.4] (box_tx) {};
        \node[anchor=center] at (box_tx.center) {Tx RF $n$};
        
        \node[fit={($(rx_digital)+(0.0,-1.7)$) ($(ant_rx_port)+(0.3,1.7)$)},fill=lightgray,behind path,scale=1.4] (box_rx) {};
        \node[anchor=center] at (box_rx.center) {Rx RF $m$};
    \end{pgfonlayer}

    \coordinate [label=right:{$g^{(m,n)}$}] (g_pos) at ($(tx_anchor_g_i)!0.5!(rx_anchor_g_q)+(1.5,0.0)$) {};

    \draw[dashed] (tx_anchor_g_i) to [out=0, in=-90, looseness=0.75] (g_pos);
    \draw[dashed] (tx_anchor_g_q) to [out=0, in=-90, looseness=0.75] (g_pos);
    \draw[->,dashed] (g_pos) to [out=90, in=0, looseness=0.75] (rx_anchor_g_i);
    \draw[->,dashed] (g_pos) to [out=90, in=0, looseness=1.0] (rx_anchor_g_q);
    
    

\end{tikzpicture}
    \vspace{.25cm}
	\caption{System model}
	\label{fig:sys_quantization}
\end{figure}%
Fig.~\ref{fig:sys_quantization} visualizes the considered system consisting of a set of $M$ \ac{IQ} receive \ac{RF} chains and $N$ \ac{IQ} transmit \ac{RF} chains, feeding to and from closely co-located dedicated transmit- and receive-antennas, respectively.
Each individual transmit chain is fed with a sequence
\begin{equation}
	\WW \in \mathbb{C}^{N \times K} \quad \text{with} \quad w^{(n,k)} \sim \mathcal{CU}(-\Vmax,\Vmax)
\end{equation}
of $K$ complex-valued time-discrete digital samples. The $n\mathrm{-th}$ row of $\WW$ represents the input of a pair of \acp{DAC} driving the \ac{RF} front-end of the $n\mathrm{-th}$ transmitting antenna.
For the remainder of the following discussions, the distribution of each sample in $\mathbf{W}$ is assumed to be independently drawn from a complex uniform distribution spanning the whole quantization range; however, all the derivations can easily be updated with alternative distribution assumptions.
After digital-to-analog conversion, the quantized value-discrete base-band signal
\begin{equation}
    \XX = \widetilde{U}_{b_\mathrm{Tx}}(\WW) = \WW + \mathbf{E}_\XX \in \widetilde{\mathbb{U}}_{b_{\mathrm{Tx}}}^{N \times K}
\end{equation}
emerges from the \acp{DAC}, where the signal variance is given by
\begin{align}
    \begin{split}
        \E{\XX\XX^\HH}
        &= \E{\WW\WW^\HH} + \E{\mathbf{E}_\XX \mathbf{E}^\HH_\XX} \\
        &= \frac{K}{6} (\Vmax^2 + \lsbtx^2) \II_M \; \text{.}
    \end{split}
\end{align}
Both the transmitting and receiving front-ends are considered to be perfectly linear, which allows for the expression of the noise-free base-band signal impinging onto the receiving \acp{ADC} corresponding to the $m\mathrm{-th}$ antenna as a weighted sum
\begin{equation}
    \tilde{y}^{(m,k)} = \sum_{n=1}^{N} g^{(m,n)} x^{(n,k)} + \sum_{n=1}^{N}\sum_{\ell=1}^{L} h^{(\ell,m,n,k)} \ast x^{(n,k)}
\end{equation}
of all transmitted signals, with $g^{(m,n)} \in \mathbb{C}$ representing the mutual coupling weights between the $n\mathrm{-th}$ transmitting \ac{RF} and receiving $m\mathrm{-th}$ \ac{RF} chain and $h^{(\ell,m,n,k)} \in \mathbb{C}$ representing $L$ environmental clutter paths reflecting transmitted waveforms back to receiving antennas.
The combination of leakage and clutter represents the self-interference channel typically suppressed in full-duplex communication systems, with another interpretation of leakage being the self-interference channel's zero-delay component.
Since monostatic sensing applications only require suppression of the leakage, the system model will assume $K=0$ from here on out, focusing on the investigation of said component, and resulting in
\begin{equation}
    \tilde{y}^{(m,k)} = \sum_{n=1}^{N} g^{(m,n)} x^{(n,k)} = \mathbf{g}^{(m)\intercal} \mathbf{x}^{(k)}
\end{equation}
as a simplified base-band system model depending only on the self-interference channel
\begin{equation}
    \GG = [\mathbf{g}^{(1)},\mathbf{g}^{(2)},\dots,\mathbf{g}^{(M)}]^\intercal \in \mathbb{C}^{M \times N} \; \text{.}
\end{equation}
If all receiving gain stages are properly configured, so that none of the \acp{ADC} are driven into their clipping ranges by leakage, the power of the base-band signals' respective leakage components
\begin{equation}
    \lVert\tilde{y}^{(m,k)}\rVert^2 = \lVert \mathbf{g}^{(m)\intercal} \mathbf{x}^{(k)} \rVert^2 \leq \lVert\mathbf{g}^{(m)}\rVert^2 \lVert\mathbf{x}^{(k)}\rVert^2 \leq 2\Vmax^2
\end{equation}
impinging onto the receiving \acp{ADC} is limited by the \acp{ADC}' acceptable input range $\Vmax$.
Imposing this condition is synonymous with a power constraint
\begin{equation}
    \lVert\mathbf{g}^{(m)}\rVert^2 \leq \frac{2\Vmax^2}{\lVert\mathbf{x}^{(k)}\rVert^2} \geq \frac{1}{N} \quad \Rightarrow \quad  \lVert\mathbf{g}^{(m)}\rVert \leq \frac{1}{\sqrt{N}}
\end{equation}
on the leakage channel's rows, considering that the \acp{DAC}' maximum output is limited by their quantization set, i.e. $|x^{(n,k)}| \leq \Vmax$.
Characterizations of the self-interference channel \cite{2012.duarte,venkatasubramanian.2019} suggest that the direct leakage channel between two antennas follows a Ricean distribution, matching the measurements in \cite{gupta.1983} showing that the mutual coupling magnitude scales with the inverse distance.
To generate a heuristic random channel roughly following measurements and adhering to the non-clipping assumptions, every channel weight
\begin{equation}
    \tilde{g}^{(m,n)}_\mathrm{Re}\!\sim\!\mathcal{N}\left( \frac{\cos \phi^{(m,n)} } {d^{(m,n)}}, 1 \right),\quad \tilde{g}^{(m,n)}_\mathrm{Im}\!\sim\!\mathcal{N}\left( \frac{\sin \phi^{(m,n)} } {d^{(m,n)}}, 1 \right)
\label{eq:g_weights}
\end{equation}
is drawn from a Ricean distribution with assumed antenna distance in wavelengths $d^{(m,n)}$ and random phase $\phi^{(m,n)}$ drawn from uniform distributions, so that
\begin{equation}
    d^{(m,n)} \sim \mathcal{U}(1, 2) \quad \text{and} \quad \phi^{(m,n)} \sim \mathcal{U}(0, 2\pi) \; \text{,}
\end{equation}
respectively.
The resulting channel rows' powers
\begin{equation}
    \mathbf{g}^{(m)} = \frac{1}{N} \frac{\mathbf{\tilde{g}}^{(m)}}{\lVert \mathbf{\tilde{g}}^{(m)} \rVert}
    \label{eq:columns}
\end{equation}
are then scaled to match the non-clipping assumption.
Introducing circularly-symmetric complex additive Gaussian noise $\mathbf{N} \in \mathbb{C}^{M \times K}$ of variance $\sigma^2$  with
\begin{equation}
    \quad n^{(m,k)} \sim \mathcal{CN}(0, \sigma^2) \quad \text{and} \quad \E{\NN^\HH \NN} = M \sigma^2 \II_K
    \label{eq:n}
\end{equation}
to the base-band leakage yields the linear model of the received base-band leakage signal
\begin{equation}
    \YY = \GG \XX + \NN = \GG (\WW + \EX) + \NN \in \mathbb{C}^{M \times K} \text{ .}
\end{equation}
Each individual receive chain is terminated by two $b_\mathrm{Rx}\text{-bit}$ \acp{ADC}
converting analog received \ac{IQ} base-band signals back to digital samples, leading to
\begin{align}
    \ZZ &= \widetilde{U}_{b_\mathrm{Rx}}(\YY) = \GG (\WW + \EX) + \NN + \EZ \in \widetilde{\mathbb{U}}_{b_{\mathrm{Rx}}}^{M \times K}
    \label{eq:Z}
\end{align}
as the final linear model describing the system in terms of its digital inputs $\WW$ and digital outputs $\ZZ$.
\section{Quantization Limitations}\label{sec:quantization}
Minimizing self-interference via digital spatial precoding is synonymous with finding a set of precoding weights $\Esp \in \mathbb{C}^{N}$ lying in, or close to, the nullspace of the self-interference channel $\GG$.
However, in practice, usually only an estimate $\EsG$ of $\GG$ is available, such that the search for $\Esp$ has to be performed on the imperfect estimate $\EsG$ instead.
The channel estimation's relationship with the underlying true channel $\GG$ to be estimated may be expressed by introducing an additive estimation error term
\begin{equation}
	\EG = \EsG - \GG \in \mathbb{C}^{M \times N} \; \text{,}
\end{equation}
the statistics of which depend on the specific estimator.
Assuming an unbiased estimator, the error's expectation
\begin{equation}
	\E{\EG} = \E{\EsG - \GG} = \boldsymbol{0}
\end{equation}
is zero, meaning the estimator's expectation is equal to the leakage channel's true weights to be estimated.
Since it is relevant to the following considerations, the error covariance accumulating at the receiving antenna streams depends on the expectation\looseness-1
\begin{equation}
	\E{\EG^\HH \EG}
	= \E{ (\EsG - \GG)^\HH (\EsG - \GG) }
	= \E{ \EsG^\HH \EsG } - \GG^\HH \GG
    \label{eq:estcov}
\end{equation}
of the channel estimation error's Gramian.
Given a vector of precoding candidate weights $\Esp$, a measure of the expected leakage power can be defined as
\begin{equation}
	\rho =  \E{ \lVert \EsG \Esp \rVert_2^2 } = \Esp^\HH \EsG^\HH \EsG \Esp
\end{equation}
the sum of squares of the channel propagation after precoding.
An ideal precoder would therefore result in $\rho = 0$, meaning no power is expected to leak from the transmitting \acp{DAC} to the receiving \acp{ADC}.
Practical precoders are required to optimize for multiple system performance metrics and usually suppress the leakage below a certain threshold level \cite{hernangomez.2026}, resulting in a non-zero expected residual leakage $\rho > 0$.
In contrast to the self-interference power's expectation based on channel estimations and unquantized precoding weights, the true self-interference power, assuming the linear system model introduced in Section \ref{sec:sys}, depends on the unknown true channel $\GG$ and the quantized transmit symbols $\XX$.
Replacing the available estimate $\EsG$ and digital precoding $\Esp$ by their additive error terms, so that
\begin{align}
    \begin{split}
	   \lVert \GG \mathbf{x} \rVert_2^2 &= \lVert ( \EsG  - \EG ) ( \Esp - \EX ) \rVert_2^2
    \end{split}
\end{align}
leads to a corrected expectation of the self-interference power
\begin{align}
    \E{ \lVert \GG \mathbf{x} \rVert_2^2 } &= \E{\Esp^\HH \EsG^\HH \EsG \Esp} + \Tr{ \Esp\Esp^\HH \E{\EG^\HH\EG} } \nonumber \\
    & \qquad+ \Tr{\E{\EX^\HH\EX} ( \EsG^\HH \EsG + \E{\EG^\HH \EG} ) } \nonumber \\
    \begin{split}
        &= \rho + \lVert \Esp \rVert_2^2 \Tr{\E{\EG^\HH\EG}} \\
        & \qquad + \frac{\lsbtx^2}{6} \ \left(\lVert \EsG \rVert_\mathsf{F}^2 +  \Tr{\E{\EG^\HH\EG}} \right)
    \end{split}
\end{align}
depending purely on the initial expected leakage, the precoding power $\lVert \Esp \rVert_2^2$, the leakage channel estimate's power $\lVert \EsG \rVert_\mathsf{F}^2$, the channel estimator's error variance $\E{\EG^\HH\EG}$, and the \acp{DAC}' parameters. 
The least-squares estimator $\EsGLS$ of $\GG$ can be defined as the received samples multiplied by the left-sided pseudo-inverse of the known quantized transmitted samples \cite[Eq. 4]{yin.2013}
\begin{equation}
    \begin{split}
    	\EsGLS = \argmin_{\EsG} \lVert \GG - \EsG \rVert_{\mathsf{F}}^2 = \ZZ \XX^\HH(\XX\XX^\HH)^{-1} \; \text{.}
    \end{split}\label{eq:EsGLS}
\end{equation}
Solving \eqref{eq:estcov} for the least-squares \eqref{eq:EsGLS} estimator by replacing $\ZZ$ with the linear system model \eqref{eq:Z} leads to
 \begin{equation}
    \E{ \EGLS^\HH \EGLS  } = \frac{M}{K} \frac{6\sigma^2 + \lsbrx^2}{\Vmax^2 + \lsbtx^2}\II_N
\end{equation}
characterizing the least-square estimator's error variance depending on the linear system model's parameters.
In combination with the nullspace precoding proposed in \cite{everett.2016} with $\rho=0$, this leads to an expected leakage
\begin{multline}
	\E{ \lVert \GG \mathbf{x} \rVert_2^2 } = \\
    \lVert \EsG \rVert_\mathsf{F}^2 \frac{\lsbtx^2}{6} + \left( \frac{\lVert \Esp \rVert_2^2 }{N} + \frac{\lsbtx^2}{6} \right) \frac{MN}{K} \frac{6\sigma^2 + \lsbrx^2}{\Vmax^2 + \lsbtx^2}
    \label{eq:theory_scenario}
\end{multline}
for a self-calibrating monostatic sensing system.
\section{Simulation}\label{sec:simulation}
The system discussed in the previous section is implemented as a Monte Carlo simulation in Python, extending the core functionalities of the \ac{HER} \cite{adler.2022} by a custom scenario.
Each experiment generates and evaluates a sequence of $10^5$ simulation samples, with each sample generating an independent channel realization via \eqref{eq:g_weights}-\eqref{eq:columns}, estimating the channel via the least-squares estimator \eqref{eq:EsGLS}, transmitting random symbols and adding random noise \eqref{eq:n}, computing a leakage-suppressing precoding via \cite{everett.2016}, and evaluating the received residual power, once again adding random noise \eqref{eq:n}.
The expectation of the received power after leakage suppression is then estimated by computing the mean over all collected simulation samples and compared with the derived theoretical expectation \eqref{eq:theory_scenario}.
The first two experiments sweep over the noise variance $\sigma^2$ and the number of transmit and receive bits, respectively.
The results are visualized in Fig.~\ref{fig:simulation_tx_quantization} and Fig.~\ref{fig:simulation_rx_quantization} and show a near-perfect fit between numerical residual leakage power and expected theoretical residual leakage power.
The sweep over the number of receive bits in Fig.~\ref{fig:simulation_rx_quantization} additionally highlights the lower bound of the transmit quantization noise, leading to a diminishing return with an increased number of receive bits.
The third experiment maintains the number of bits and varies the number of observations $K$ used for leakage channel estimation and the noise power instead, highlighting the fact that the theoretical model does not accurately capture clipping effects due to noise, leading to a divergence between theory and numerical results in the domain with few observations.
\begin{figure}
	\centering
	\input{plots/simulation_tx_quantization_leakage.pgf}
	\caption{Simulated influence of transmit quantization on transmit-receive leakage with $K = 1000$ observations and $b_\mathrm{Rx}= \SI{8}{\bit}$ receive quantization}
	\label{fig:simulation_tx_quantization}
    \vspace*{.25cm}
	\input{plots/simulation_rx_quantization_leakage.pgf}
	\caption{Simulated influence of receive quantization on transmit-receive leakage with $K = 1000$ observations and $b_\mathrm{Tx}= \SI{8}{\bit}$ transmit quantization}
	\label{fig:simulation_rx_quantization}
    \vspace*{.25cm}
	\input{plots/simulation_num_observations_leakage.pgf}
    \caption{Simulated influence of number of observations on transmit-receive leakage with $b_\mathrm{Rx}= \SI{8}{\bit}$ receive quantization and $b_\mathrm{Tx}= \SI{8}{\bit}$ transmit quantization}
	\label{fig:simulation_num_observations}
    \vspace*{-.5cm}
\end{figure}
\section{Measurements}\label{sec:measurements}
\begin{figure}
	\centering
	\input{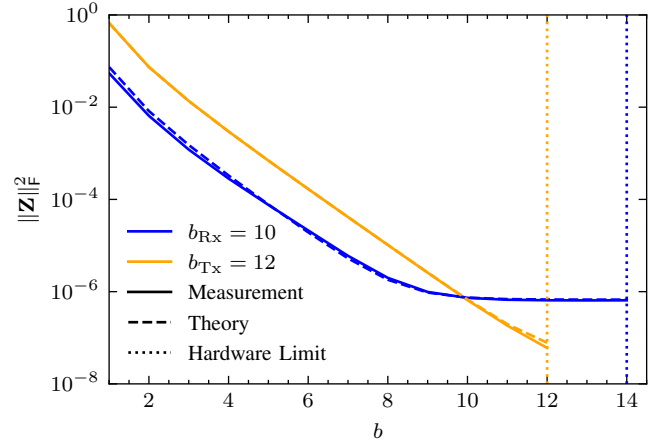}
	\caption{Measured influence of number of transmit- and receive bits on transmit-receive leakage in a $3\!\times\!3$ \ac{MIMO} testbed with $K = 1000$ observations}
	\label{fig:measurement_quantization}
\end{figure}

In order to demonstrate and validate the theoretical and numerical investigations in the previous section, the least-squares leakage estimation \eqref{eq:EsGLS} and successive leakage cancellation precoding \cite{everett.2016} are deployed to an \ac{SDR} testbed consisting of a single Ettus X410 \ac{USRP}
driving an array of three transmitting and three receiving horn-antennas.
The \ac{SDR} features transmit \acp{DAC} with $b_\mathrm{Tx,SDR} = \SI{14}{\bit}$ and receive \acp{ADC} with $b_\mathrm{Rx,SDR} = \SI{12}{\bit}$, respectively, both resolving maximum amplitude of $V_\mathrm{Max} = 1$.
Since the hardware's number of quantization bits is naturally immutable, the effect of a decreased amount of quantization bits can only be mocked in real systems by digitally introducing additional uniformly distributed noise
\begin{equation}
    n_\mathrm{Tx,SDR}^{(n,k)},\ n_\mathrm{Rx,SDR}^{(m,k)} \sim \mathcal{CU}(-1,1)
\end{equation}
to the transmitted and received digital samples.
It is scaled to correct the overall introduced transmit and receive quantization noise power, so that
\begin{align}
	w_{b_{\mathrm{Tx}}}^{(n,k)} &= w^{(n,k)} + \sqrt{\lsbtx^2 - \Delta_{b_\mathrm{Tx,SDR}}^2} n_\mathrm{Tx,SDR}^{(n,k)}\\
    z_{b_{\mathrm{Rx}}}^{(m,k)} &= z^{(m,k)} + \sqrt{\lsbtx^2 - \Delta_{b_\mathrm{Rx,SDR}}^2} n_\mathrm{Rx,SDR}^{(m,k)}
\end{align}
are the updated digital samples uploaded to and downloaded from the \ac{USRP}.
The \ac{USRP}'s master clock rate and firmware image is configured to result in a \ac{DAC} and \ac{ADC} sampling rate of $\SI{4}{\MHz}$, with an additional digital Butterworth lowpass filter decimating the effective measurement bandwidth by a factor $32$ to $\SI{125}{\kHz}$ in order to ensure adherence to the system model's flat channel assumptions.
The device's front-ends are configured to a carrier frequency of $\SI{6}{\GHz}$, with the amplifiers providing $\SI{30}{\dB}$ receive amplification and $\SI{50}{\dB}$ transmit amplification.
Two measurement sets are generated, the first sweeping over the number of transmit bits while holding the number of receive bits steady, the second sweeping over the number of receive bits while maintaining the number of transmit bits.
The results are visualized in Fig.~\ref{fig:measurement_quantization}, exhibiting a good fit to the theory derived in Section \ref{sec:quantization}.
\section{Conclusion}\label{sec:conclusion}
This work investigated the impact of transmit- and receive-quantization noise on joint \ac{MIMO} leakage estimation and subsequent suppression through digital spatial precoding approaches.
Based on a simplified linear system model, a closed-form term for the expected residual leakage power was derived and compared to numerical simulations as well as real-world measurement data.
The results indicate that it is possible to accurately predict the performance by correcting the precoding's expected residual leaking power with a term depending only on known system parameters, enabling more precise leakage-suppression algorithm development before hardware implementation.
While this is a relevant interim result, future investigations should additionally consider the impact of clock jitter, phase noise, amplification nonlinearities, and frequency-selectivity of the leakage channel, requiring more complex system models.
\bibliography{references.bib}
\end{document}